# A Miniaturized Low Frequency (LF) Magnetoelectric Receiving Antenna with an Integrated DC Magnetic Bias

Yunping Niu[1,2,3] and Hao Ren[1,*]

[1] School of Information Science and Technology, ShanghaiTech University, Shanghai, 201210, China
[2] Shanghai Institute of Microsystem and Information Technology, Chinese Academy of Sciences, Shanghai, 200050, China.
[3] University of Chinese Academy of Sciences, Beijing, 100049, China.
*E-mail: renhao@shanghaitech.edu.cn

## Abstract

Due to the applications in meteorological broadcasts, radio navigation and underwater communications, low-frequency(LF) receiving antennas have been actively studied. However, because the frequency range of LF antenna is 30kHz to 300kHz, its electromagnetic wavelength is 1km to 10km, which makes LF electromagnetic antennas difficult to be implemented in miniaturized or portable devices. This article presents a miniaturized LF magnetoelectric(ME) receiving antenna with an integrated DC magnetic bias. The antenna is based on the magnetoelectric effect and operates by resonance at its mechanical resonant frequency. Thus, compared with traditional LF wire antennas, the dimension of ME antenna is reduced significantly. Compared with prior art of ME antennas which do not have DC magnetic bias, higher performance can be achieved by integrating the miniaturized DC magnetic bias. Compared with prior art of an ME antenna with bulky external DC magnetic bias, the ME antenna with an integrated DC magnetic bias significantly reduce its dimension. Magnetostrictive $TbDyFe_2$(Terfenol-D) and piezoelectric lead zirconate titanate(PZT) thin films are bonded together to form the 38×12×5.8mm³ ME receiving antenna. Four 10×10×10mm³ Rb magnets are implemented to provide an optimal DC bias for the antenna. A maximum operation distance of 2.5m is demonstrated with the DC magnetic field bias, 2.27 times of the maximum operation distance of the antenna without DC magnet field bias. The efficiency, gain and quality factor the ME receiving antenna is also characterized. The miniaturized LF ME antenna could have potential applications in portable electronics, internet of things and underwater communications.

Low frequency (LF) antennas, which coverts LF electromagnetic wave into voltage or converts voltage into electromagnetic wave, have been widely implemented in meteorological broadcasts, radio navigation, radio frequency identification (RFID) and underwater communication[1-7]. LF is generally designated as electromagnetic waves with frequencies between 30kHz-300kHz, which corresponds to wavelengths from 1km to 10km[8]. In RFID applications, the LF signal not only can transfer signals, it also can be implemented for wireless power transfer[2], corrosion detection[3, 4] and underground object detection of rocks and soil[5, 6]. For underwater communication, the RF communication distance is significantly reduced because dielectric loss in water increases with frequency and the electromagnetic waves is attenuated significantly at high frequency[7]. As a result, implementing low-frequency electromagnetic waves mitigates the dielectric loss. Aforementioned studies have shown that LF band has useful applications such as communications and sensing, however, conventional LF antennas are large in dimension because they are directly driven by electric current or voltage to accelerate the electrons inside the metal plates for radiation, which requires the antenna dimension to be comparable to the wavelength [9, 10]. For instance, the 60kHz WWVB antenna build by National Institute of Standards and Technology (NIST) is large in dimension and is supported by 122m-high towers[11]. The large dimension makes it difficult to be implemented in communications and sensing for portable devices[12]. To reduce the dimension of LF antennas for portable devices, several LF antenna structures have been proposed. James et al. presented a LF mechanical antenna based on electret in motion. Since a rotating electret would appear as a radiating electric dipole, the electromagnetic wave can be transmitted in mechanical motion[13, 14]. Cao et al. presented a LF transmitting antenna based on spinning magnet. By moving the permanent magnet mechanically, the magnetic field in the space changes in time, thus the electromagnetic field is transmitted[15].

In addition to previous mentioned studies, magnetoelectric effect is also a promising technique for antenna miniaturization. In the past few years, several studies have documented antennas based on the magnetoelectric(ME) effect[10, 16-21]. These ME antennas receive magnetic waves through the ME effect at their mechanical resonance frequencies. The ME effect in a material is an electric polarization response to an applied magnetic field, or a magnetization response to an applied electric field[22, 23]. The ME effect is normally realized by bonding magnetostrictive and piezoelectric materials. The magnetostrictive material generates a strain under a magnetic field, which drives the piezoelectric material to produce electric charge, thus interconnection between magnetism and electricity is realized. Currently the commonly implemented magnetostrictive materials include $TbDyFe_2$(Terfenol-D), $Ni_{0.8}Zn_{0.2}Fe_2O_4$, FeGaB and metglas[21, 24-26]. The piezoelectric materials include PZT and AlN[17, 21, 25]. Prior studies in magnetoelectric energy converters have demonstrated that the ME coefficient is related to the DC magnetic field bias. As the magnetic field increases, the strain in magnetostrictive layer increases as well. However, the slope of increase is not constant: it first rises and then falls[27]. This characteristic means that the ME coefficient reaches a maximum at an optimal magnetic field bias[28-31]. Many ME energy converters or gyrators have adopted bulky electromagnets or Helmholtz coils to provide magnetic bias to enhance the output voltage. Xu et al. have implemented a Helmholtz coil to improve the performance of ME antenna, however, it significantly increases the dimension[32]. To date, no miniaturized DC magnetic field bias have been applied in ME antennas to improve the performance. Besides, ME antennas operating at VLF[16], HF[17], VHF[19, 20] and UHF[18, 21] have been reported, however, no ME antenna at LF band has been reported. In addition, magnetic materials are not compatible with electromagnetic interference(EMI) sensitive electronics, such as inductors and power MOSFET[33], and thus many transmitting antennas may not be made by ME materials. Currently most state of art RFID applications implement inductive coils as transmitters, due to the low cost and ease of fabrication. When alternating current flows through an inductive coil, it generates alternating magnetic field. Currently no prior studies have demonstrated ME antennas to receive magnetic field from an inductive coil.

In this paper, we report a miniaturized LF magnetoelectric receiving antenna with an integrated DC magnetic bias. Four Rb magnets are implemented to provide optimal DC magnetic bias, which results in extended operation range. By adopting Rb magnets to provide DC bias magnetic field rather than bulky electromagnets or Helmholtz coils, a miniaturized ME antenna with a dimension of 38×12×5.8mm$^3$ is demonstrated with a maximum detection range of 2.5m when a 40-turn spiral coil with an outer diameter of 50 mm is implemented as the transmitter.

Figure 1(a) and (b) present the schematic and optical image of the LF receiving antenna and testing platform. The receiving antenna has a sandwich structure of one PZT layer and two Terfenol-D layers. The PZT layer is a stack of 14 pieces of 0.2mm thick PZT laminate and the total thickness is 3.4mm. The PZT layer is glued in the middle of two Terfenol-D layers by epoxy glue. The dimensions of PZT and Terfenol-D layers are $38\times12\times3.4mm^3$ and $35\times8\times1.2mm^3$. Three additional devices with different PZT thickness (0.25mm, 0.8mm and 1.5mm) are fabricated. The two surfaces of PZT are coated with silver paste as electrodes and two wires are connected to the silver paste on the two sides as interconnect. A 40-circle spiral coil milled on a FR4 PCB board is implemented as the transmitting antenna. The outer diameter of the spiral coil is 50mm, and the line width and gap of each turn is 0.25mm. Four $10\times10\times10mm^3$ Rb magnets are placed on the receiving antenna to provide a DC magnetic field bias for receiving antenna to improve its performance.

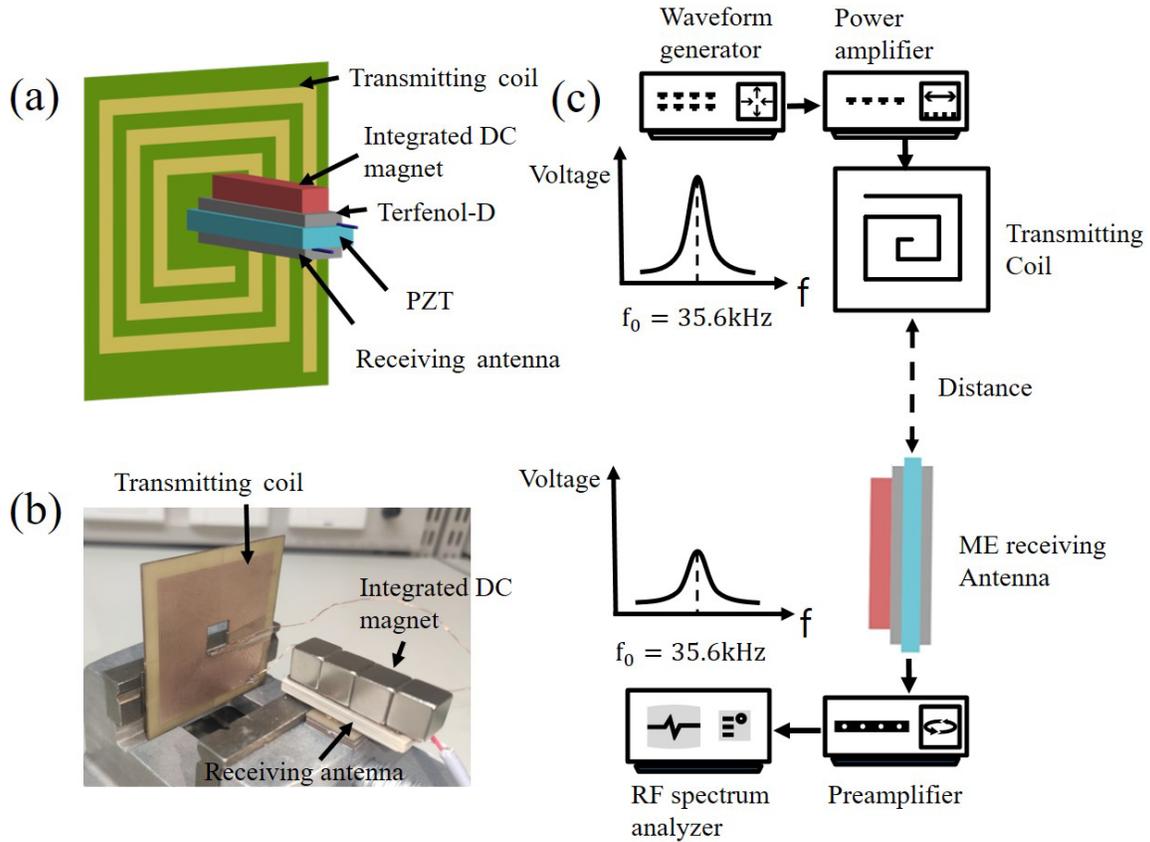

**Figure 1(a)The schematic of the ME receiving antenna; (b)The optical image of the ME receiving antenna; (c)The schematic of testing platform.**

$TbDyFe_2$(Terfenol-D) is a giant-magnetostrictive-effect material[29, 31, 34, 35]. Upon a change of magnetic field, its dimension changes due to magnetization. When operating as a receiving antenna, an electromagnetic wave in the space drives the Terfenol-D laminate to contract/extend. The PZT layer bonded between the Terfenol-D layers vibrates synchronously. Due to the piezoelectric effect, the strain evokes charge accumulation at the two electrodes of the PZT and a voltage is generated. In this way the electromagnetic wave is converted to voltage and can thus be detected, and the sandwich structure operates as a receiving antenna.

The schematic of the test platform is illustrated in Figure 1(c). A sine wave signal is generated by a waveform generator(DG1062Z, RIGOL Technology Inc.) and amplified by a power amplifier(ATA-1200A, Aigtek Inc.). The amplified signal is fed to a transmitting coil to generate AC electromagnetic wave. The electromagnetic wave is received by the receiving antenna and then amplified by a low-noise preamplifier(SR560, Stanford Research Systems Inc.). The amplified signal is fed to a RF spectrum analyzer(FSW8, Rohde & Schwarz GmbH.) or an oscilloscope(MSO2024B, Tektronics Inc.) for signal recording.

To obtain the maximum output voltage, a maximum strain needs to be generated. Therefore, the operation frequency of the receiving antenna should be set at its mechanical resonant frequency in the longitudinal direction. The i-th resonant frequency in the longitudinal direction of a multi-layer laminated rectangular plate is calculated by Equation S-1. The calculated resonant frequency of our device is 37.7 kHz, in agreement with the finite element analysis results of 39.7 kHz shown in Figure S-1.

The resonant frequency of the receiving antennas is tested by an impedance analyzer (E4990A, Keysight Technology Inc.). Figure 2(a) shows the impedance and phase of the ME receiving antenna. The measurement resonant frequency is 35.6kHz which is slightly lower than equation S-1 and FEA prediction, possibly due to extra mass of the wires bonded to the PZT. The frequency response of the receiving antenna is shown in Figure 2(b). When the operation frequency is shifted 1.97% to 36.3kHz and 34.9kHz, the normalized output voltage decreases to 0.707 of maximum voltage.

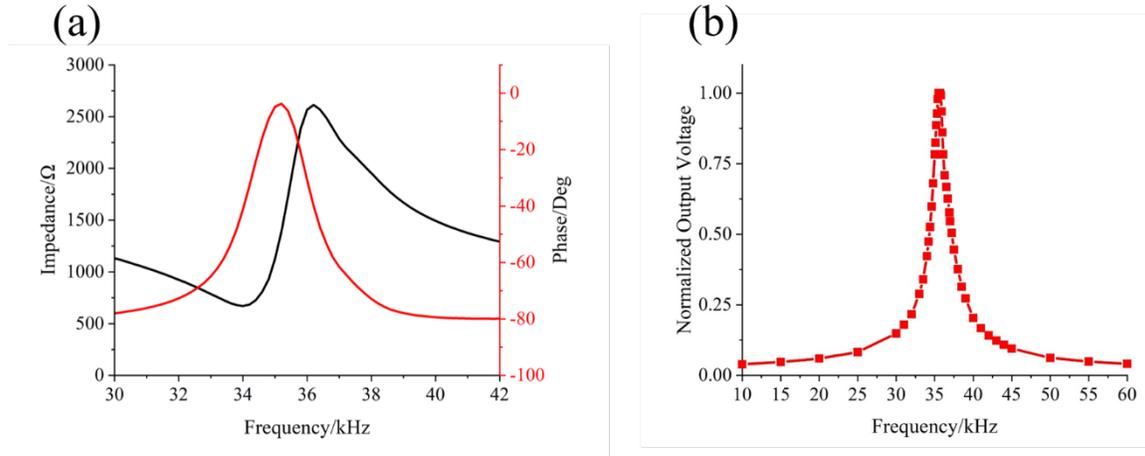

**Figure 2(a) The impedance versus frequency for different ME antennas; (b)The normalized output voltage as a function of input signal frequency.**

Four integrated Rb magnets provide DC magnetic bias to improve the performance of the receiving antenna. Previous studies demonstrated the strain in Terfenol-D layer increases with a certain DC magnetic bias[22, 23, 29, 31, 36]. Therefore, many studies have adopted bulky electromagnets or Helmholtz coils to enhance the output voltage of ME energy converters. However, it significantly increases the size of the receiving antenna. To enhance the output voltage with a small footprint, four integrated magnets are placed on the receiving antenna to provide the DC magnetic bias. The effect of magnetic bias is verified by applying a magnetic field along the longitudinal direction of receiving antenna using an electromagnet. Figure 3 illustrates the normalized output voltage as a function of applied DC magnetic field. The blue line is the experimental result without integrated magnets. The output voltage rises first then falls in both forward and reverse direction. The maximum voltage occurs at a DC magnetic field bias between -470 to -320 and 320 to 470Oe. With a gauss meter, the magnetic field generated by integrated magnets is measured as 320Oe in reverse direction, which is optimal to enhance the strain in Terfenol-D layer and improve the performance of receiving antenna. The red line is the test result with integrated magnets. The output voltage reaches a maximum at zero bias because the integrated magnet provides a -320Oe magnetic field. The magnetic field bias test agrees with previous work and demonstrates the integrated magnets provides optimal DC magnetic bias to improve the performance of the receiving antenna.

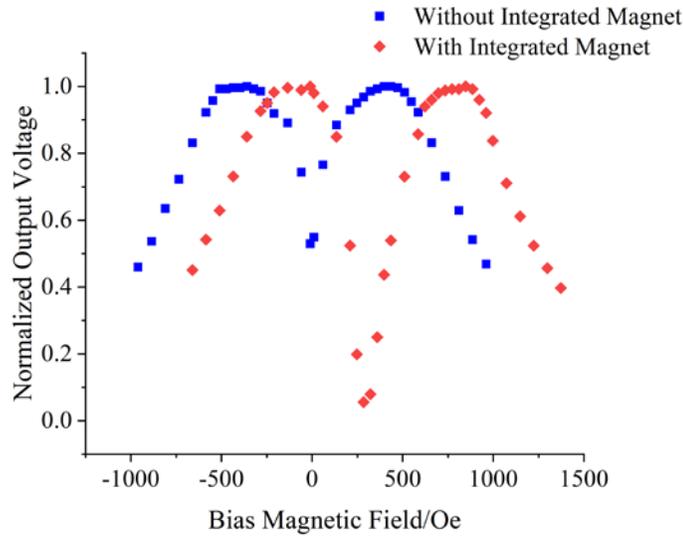

**Figure 3 Normalized output voltage as a function of magnetic field bias with and without integrated DC magnets.**

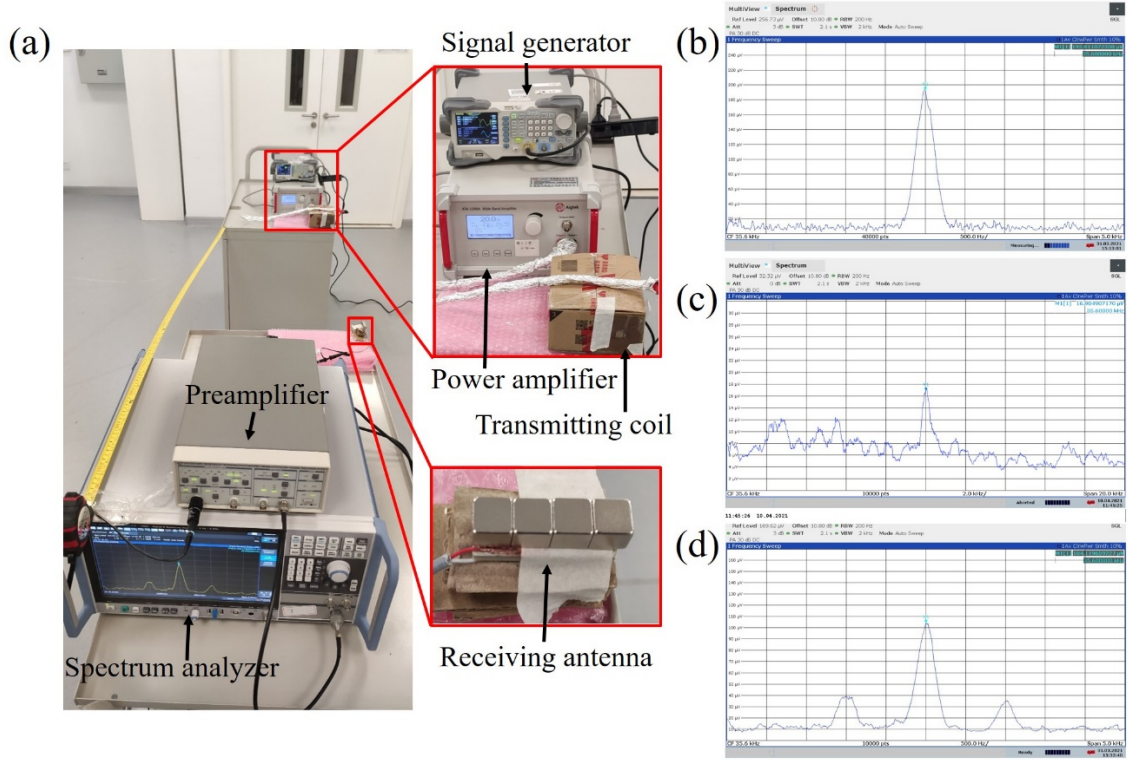

**Figure 4(a)Optical image of the ME antenna measurement setup; (b)RF spectrum of the received signal from the ME antenna at 1.1m; (c)RF spectrum of the received signal from the ME antenna at 2.5m; (d)RF spectrum of the received signal from the ME antenna at 1.1m with an 1kHz sine wave AM modulation.**

To determine the maximum operation distance of the ME antenna, the output voltage is recorded at different distances between the transmitting coil and the receiving antenna, as shown in Figure 4(a). The amplified signal has an amplitude of 30Vpp at a frequency of 35.6kHz. The impedance of the transmitting coil at

35.6kHz is $Z_{coil} = 22e^{j28°} \Omega$. The input power is $P_{in} = \frac{V_{rms}^2}{|Z_{coil}|}cos\theta = 4.5W$. The RF spectrum of the output signal of the receiving antenna at both 1.1m and 2.5m distance between the transmitting coil and receiving antenna is shown in Figure 4(b) and (c), respectively. When the distance is 2.5m, the signal to noise ratio is calculated as 6.5dB, with received signal power level of -82.43dBm and the noise floor power level of -88.93dBm. Figure 4(d) shows the spectrum of the received signal when the transmitted signal is AM modulated at 1.1 m between the transmitting coil and receiving antenna. To obtain the maximum voltage gain, the carrier wave is set at 35.6kHz. The modulation signal is a 1kHz sine wave with a modulation depth of 100%.

The output voltage of the ME antenna versus distance between the transmitting coil and the receiving ME antenna is shown in Figure 5. The output voltage decreases with the increase of distance. At a distance of 2.5m, the output voltage is 0.0169mV for the ME antenna with integrated magnets, which is reduced by $5.1 \times 10^3$ times compared with the output voltage at a distance of 10cm(87mV). Without integrated magnets, the ME antennas with a PZT thickness of 0.25mm, 0.8mm, 1.5mm and 3.4mm yield a maximum distance of 0.2 m, 0.35m, 0.8m and 1.1m, respectively. The reason for the larger maximum distance for the ME antenna with a PZT thickness of 3.4mm is the stacked PZT laminates are in series connection, which increases the output voltage. In addition, with the integrated magnets on the receiving antenna, both the output voltage and the maximum operation distance are increased. The ME antennas with a PZT thickness of 0.25mm, 0.8mm, 1.5mm and 3.4mm with Rb magnets yield a maximum distance of 0.35m, 0.7m, 1.1m and 2.5m, respectively, which is 1.75, 2, 1.375 and 2.27 times larger compared to those without integrated magnets. At the distance of 0.2m, the output voltage is increased by 1.62, 11.2, 6.19 and 19 times respectively with a PZT thickness of 0.25mm, 0.8mm, 1.5mm and 3.4mm compared with the ME antennas without integrated magnets. The experiment results demonstrate that the DC magnetic field bias generated by integrated magnets enhances the ME effect and improves the performance of the antenna. The reason that DC magnets can improve the ME effect is it generates a DC magnetic field bias, which induces a larger AC strain when an AC magnetic field is applied by the spiral coil transmitter. When the DC magnetic field increases, the strain in magnetostrictive material first increases, while after reaching a maximum value, the strain decreases. With a very small or very large magnetic field bias, the strain hardly changes under an AC magnetic field. Under an optimal bias magnetic field, the amplitude of AC strain reaches a maximum and induce a higher voltage between the two electrodes of PZT.

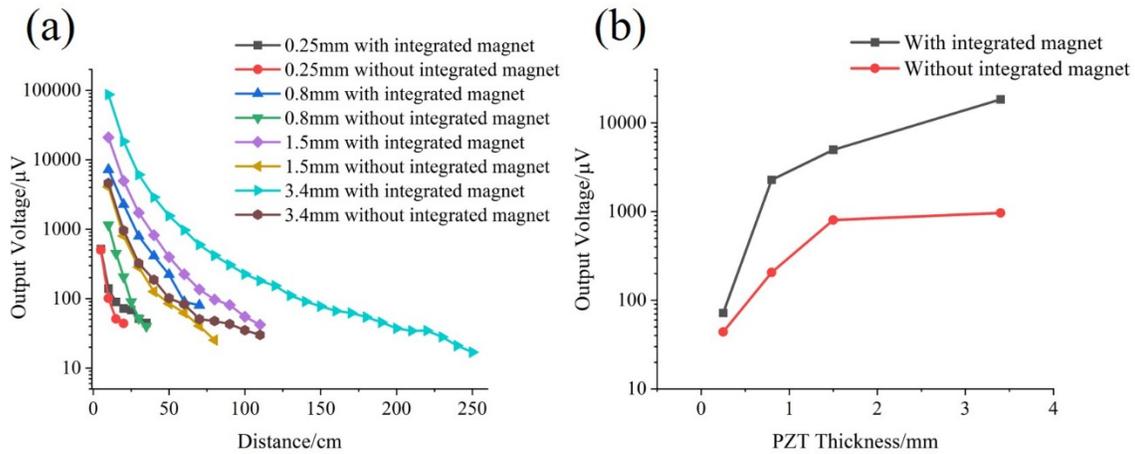

**Figure 5(a)The output voltage vs. distance plot; (b)The output voltage of different thickness PZT layers at 0.2m distance with and without integrated DC magnet.**

The angle of rotation around the short edge of the ME antenna is varied to evaluate its impact on the receiving antenna performance. Figure 6(a) and (b) show the output voltage versus the angles at a distance of 10mm and 70mm respectively. When the receiving antenna is close to the transmitting coil (10mm), the parallel direction received a 1.5-times higher voltage compared to the vertical direction. This is because at a small distance, the magnetic field generated by the transmitting coil in vertical and parallel direction both have a large value. When the receiving antenna rotates from 90 degrees to 0 and 180 degrees, the average distance between receiving antenna and transmitting antenna is reduced. Therefore, the parallel direction provides a larger voltage. When the distance increases to 70mm, the vertical direction received a 1.7-times higher voltage compared to the parallel direction. This is due to the spiral planar coil transmits vertical magnetic field effectively while the parallel magnetic field is significantly attenuated. At a greater distance, the magnetic field intensity in parallel direction is much weaker than vertical direction, which leads to a lower voltage. Figure 6(b) also shows at 70mm, the 3dB beam width is 110º. The directivity is estimated as 2.25 (3.5dB) according to the pattern plot. The impact of rotation angle around the center axis and the horizontal translation between the two antennas on the performance of the ME antenna is shown in the supplementary materials.

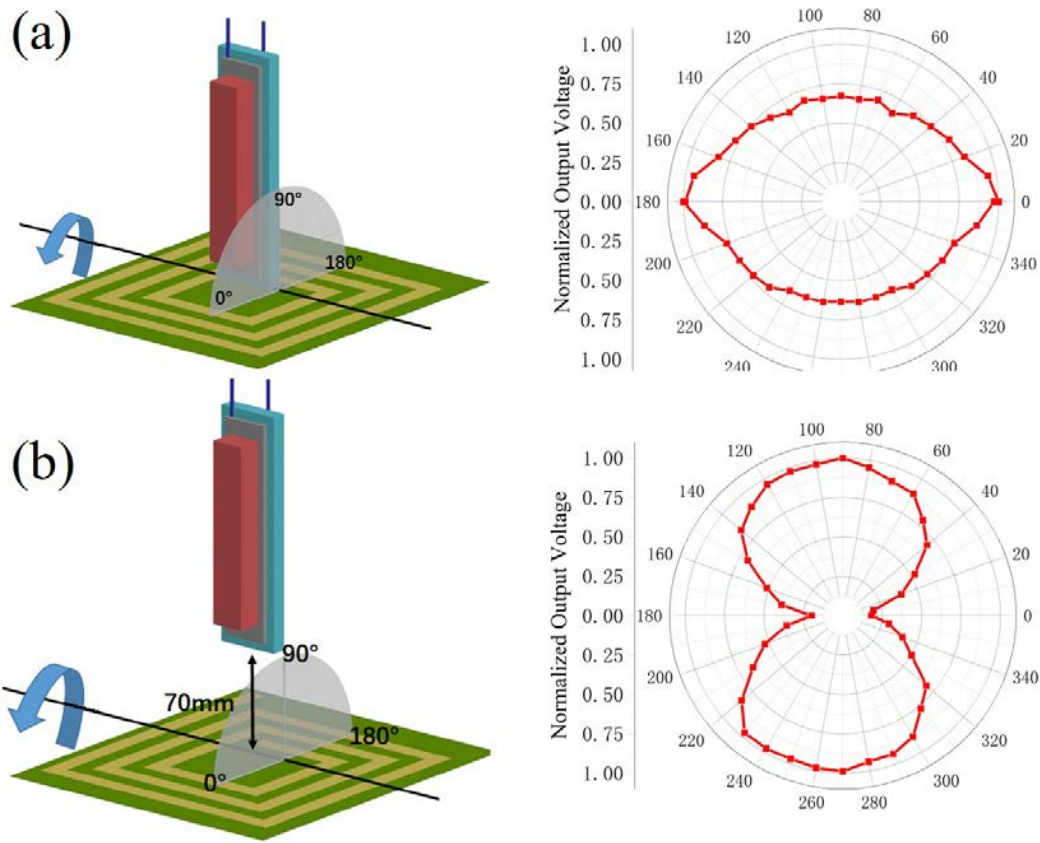

**Figure 6 Experimental result of the normalized output voltage versus rotation angle around the short edge:(a)The distance between transmitting and receiving antenna is 10mm; (b)The distance between transmitting and receiving antenna is 70mm.**

The efficiency and gain of the ME receiving antenna is also characterized. In order to measure the efficiency, we connect it to a waveform generator to set it as a transmitting antenna because an antenna has the same efficiencies and gains when implemented as transmitting and receiving antenna[37]. Due to the long wavelength of kHz electromagnetic wave, it is difficult to measure the transmitted power of this ME antenna. As a result, the efficiency of the ME antenna can be inferred by comparing with a theoretical circular loop antenna[32]. The near field magnetic flux density of the small circular loop antenna can be derived by equation S-2. Figure 7 shows the measured magnetic flux density of the ME antenna and the analytical magnetic flux density of the small circular loop antenna as a function of distance in two directions, i.e., $\theta = 90º$ and $\theta = 0º$. The measurement setup of magnetic flux density is shown in the supplementary material.

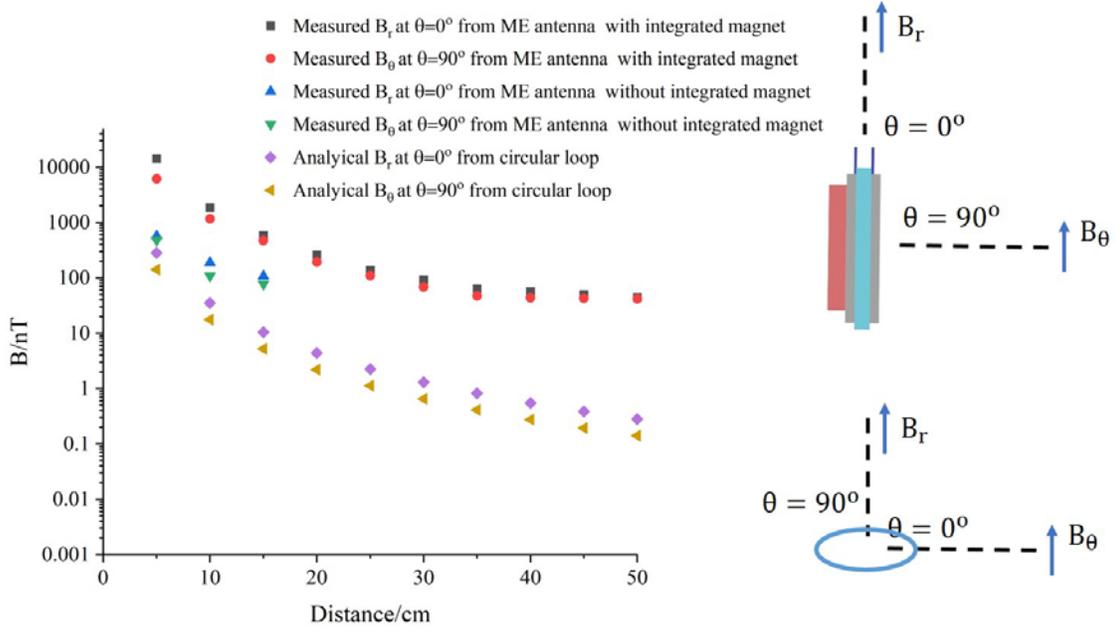

**Figure 7 The measured and analytical magnetic flux density of ME antenna and small circular loop antenna.**

The magnetic flux density of the ME antenna in Figure 7 is approximately 50 times higher than the circular loop. By calculating the radiation efficiency of the circular loop antenna, the radiation efficiency of the ME antenna can be acquired. The radiation efficiency is proportional to the squared magnetic flux[32]. Therefore, the efficiency ratio of ME antenna and circular loop is derived as:

$$\frac{\eta_{ME}}{\eta_{loop}} = (\frac{|B_{ME}|}{|B_{loop}|})^2 \approx 2.5 \times 10^3 \qquad (2)$$

The radiation efficiency of a circular antenna is derived as[32, 38]:

$$\eta_{loop} = \frac{R_r}{R_r + R} \approx \frac{R_r}{R} = 2.6 \times 10^{-20} \qquad (3)$$

The efficiency of the circular loop is extremely low because its size is much smaller than its wavelength and the antenna is non-resonant[39]. The ME antenna operating at mechanical resonance enhances the efficiency by 2500 times to $6.5 \times 10^{-17}$. The gain of ME antenna can be calculated as $G = \eta_{ME} D = 1.46 \times 10^{-16}$. Here $D$ is directivity of the ME antenna and D=2.25 based on Figure 6. Compared to the ME antenna without DC magnetic bias, the efficiency of the ME antenna with bias increases by 620 times. The efficiency of the ME antenna is in agreement with previously published work, which is on the order of $10^{-16}$ [32].

The quality factor of ME antenna can be calculated as follow:

$$Q_{ME} = \frac{f}{BW_{3-dB}} \qquad (4)$$

As a result, the calculated $Q_{ME}$ is 25.4, which is in agreement with previously published work of 200.2[40]. The calculation process is shown in the supplementary material.

In summary, a miniaturized LF magnetoelectric(ME) receiving antenna with an integrated DC magnetic field bias is presented. The DC magnetic field bias is achieved by implementing four Rb magnets on top of the ME antenna, and the magnetostrictive layer in the ME antenna generates larger strain when the same AC magnetic field is generated by a transmitting spiral coil, resulting in higher voltage output from the piezoelectric layer in the ME antenna as well as an extended maximum operation range. With a size of 38×12×5.8mm$^3$, the ME antenna demonstrates a maximum operation distance of 2.5m with the Rb magnet bias, 2.27 times of the maximum operation distance of the ME receiving antenna without Rb magnet bias. The LF ME receiving antenna may find potential application in miniaturized portable electronics, internet of things and underwater communications.

## Availability of data
The data that support the findings of this study are available from the corresponding author upon reasonable request.

## Declaration of interests
The authors declare that they have no known competing financial interests.

## Author contribution
H.R. conceived the idea. H.R. and Y.N. designed the experiments. Y.N. implemented the experimental set-up and conducted experiments. Y.N. and H.R. analyzed data and participated in discussions. Y.N. and H.R. wrote the paper. H.R. led the work.

# A Miniaturized Low Frequency (LF) Magnetoelectric Receiving Antenna with an Integrated DC Magnetic Bias (Supplementary Materials)

Yunping Niu[1,2,3] and Hao Ren[1,*]

[1] School of Information Science and Technology, ShanghaiTech University, Shanghai, 201210, China

[2] Shanghai Institute of Microsystem and Information Technology, Chinese Academy of Sciences, Shanghai, 200050, China.

[3] University of Chinese Academy of Sciences, Beijing, 100049, China.

*E-mail: renhao@shanghaitech.edu.cn

## 1. Mechanical resonate frequency of the magnetoelectric receiving antenna

For a composite rectangle plate, its mechanical resonate frequency in longitudinal direction can be calculated as[1]:

$$f_i = (2i-1)\frac{1}{2L}\sqrt{\frac{CA_e}{\rho A_e}}$$

$$CA_e = C_1 A_1 + C_2 A_2$$

$$\rho A_e = \rho_1 A_1 + \rho_2 A_2$$

(S-1)

Where $L$ is the length of the structure. $A$ is the cross sectional area. $CA_e$ and $\rho A_e$ represents the effective axial rigidity and equivalent mass per length of the beam, where $C$ is the Young's modulus of the material. The parameters of PZT and Terfenol-D are listed in Table S-1[2]. The calculated resonant frequency of our device is 37.7 kHz.

**Table S-1 Parameters of PZT and Terfenol-D**

|  | PZT | Terfenol-D |
| --- | --- | --- |
| Young's Module (GPa) | 82.1 | 50 |
| Density (kg/m$^3$) | 7750 | 9250 |

To verify the calculated result, the ME antenna structure is simulated in FEA software. The simulated result shows the mechanical resonate frequency in longitudinal direction is 39.7kHz as shown in Figure S-1, which is in agreement with analytical result and experiment result.

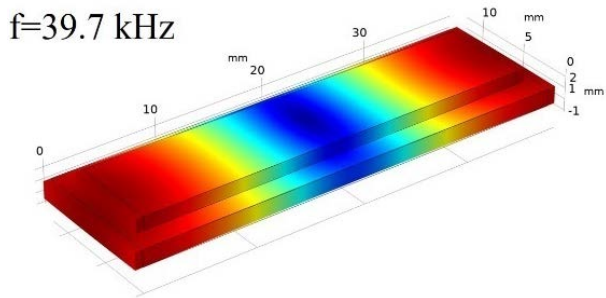

**Figure S-1 Simulation result of resonant frequency by FEA.**

## 2. Efficiency measurement

In order to measure the efficiency of proposed ME antenna, we set it as a transmitting antenna. The efficiency can be inferred by comparing with a theoretical circular loop antenna[3]. The measurement setup is shown in Figure 2. The magnitude of magnetic flux density is measured by an inductor coil, which is made of 0.06mm diameter enameled wire winding. The number of turns of the coil is 10000 and the diameter of coil is 1.5cm. According to the law of electromagnetic induction, the induced voltage of the coil is in proportional to the magnitude of alternating magnetic field. The coil is calibrated by placing the coil in a standard alternating magnetic field and recording the change in output voltage due to the standard alternating magnetic field. Afterwards, the coil can be implemented to measure an unknown alternating magnetic field by measuring the output voltage of the coil.

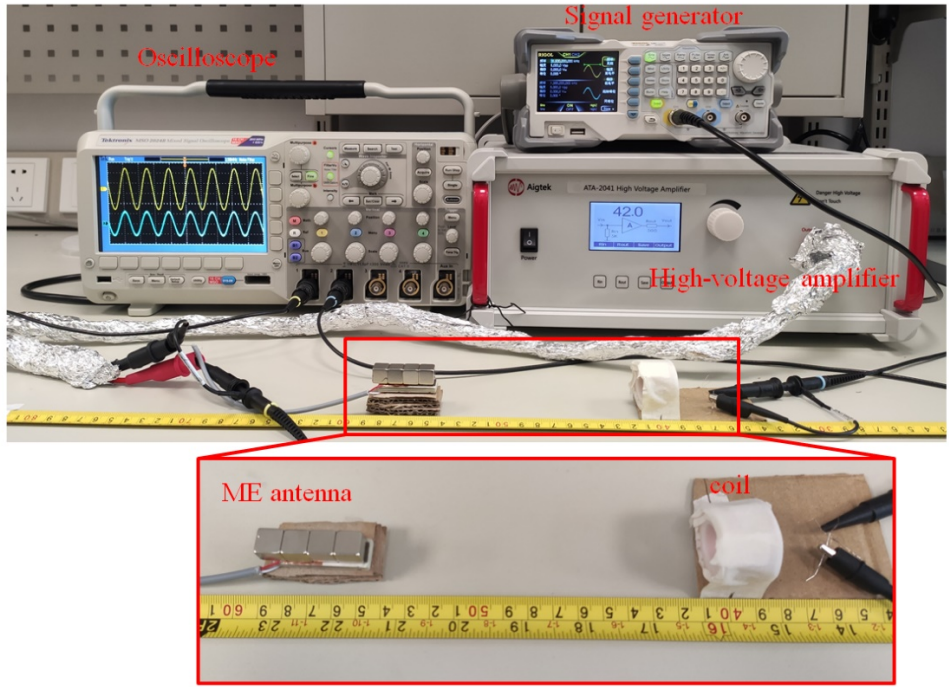

**Figure S-2 The measurement setup of magnetic flux density.**

The near field magnetic flux density of the small circular loop antenna can be derived as.

$$|B_r| \approx \frac{\mu_o}{\sqrt{2}} \sqrt{\frac{3}{\pi \eta k^4} \frac{Z_o R_r P_{in}}{(R_r + Z_o)^2}} (\frac{2\cos\theta}{r^3})$$

$$|B_\theta| \approx \frac{\mu_o}{\sqrt{2}} \sqrt{\frac{3}{\pi \eta k^4} \frac{Z_o R_r P_{in}}{(R_r + Z_o)^2}} (\frac{\sin\theta}{r^3})$$

(S-2)

where $\mu_0$ is the permeability of the free space($4\pi \times 10^{-7}$H/m), $\eta$ is the free space radiation impedance (377Ω), $k$ is the propagation constant for a wavelength $\lambda$ ($k = \frac{2\pi}{\lambda}$); $Z_0$ is the impedance of the transmission line ($Z_0 = 50\Omega$), $R_r$ is the radiation resistance of the antenna and $S$ is the area of the circular loop ($R_r = \frac{8}{3}\eta\pi^3(\frac{S}{\lambda^2})^2 \approx 31171\frac{S^2}{\lambda^4}$), $P_{in}$ is the input power which is the same power provided to the ME antenna ($P_{in} = 1.8$W).

As shown in Figure S-2, a sine wave signal is generated by a signal generator and fed to a high voltage power amplifier. The amplified signal with a frequency of 35.6kHz and an amplitude of 200V Vpp is fed to the ME antenna, causing it to vibrate and an AC magnetic field is generated. The magnetic field generated by the ME antenna is sensed by the detection coil, resulting in an output voltage at the detection coil. This output voltage is recorded by an oscilloscope. As a result, the magnitude of magnetic field generated by the ME antenna can be calculated.

The calibration data of the magnitude of standard alternating magnetic field versus the output voltage of the coil is shown in Figure S-3. According to Figure S-3, the magnitude of an unknown magnetic field can be calculated based on the voltage of the coil induced by the unknown magnetic field. The result of the linear regression is:

$$y = 2.677x - 20.5758$$

(S-3)

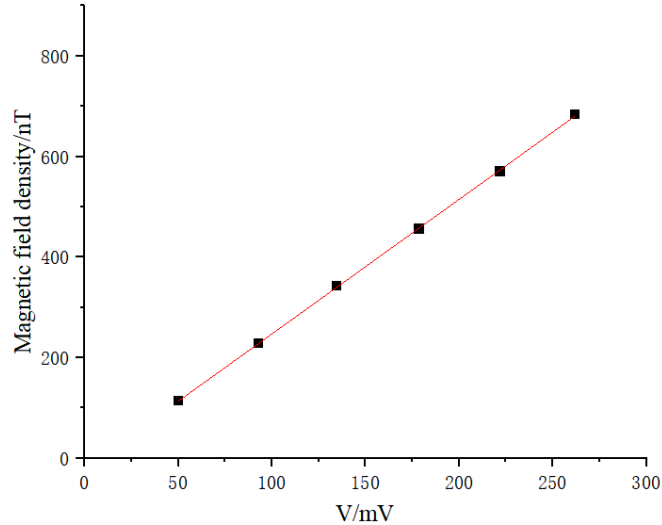

**Figure S-3 The calibration data of detection coil.**

3. **Radiation quality factor measurement**

To measure the radiation quality factor of ME antenna, it is comparable with the radiation quality factor of the same size ESA (electric small antenna)[4]. The quality factor of the ME antenna can be estimated as follows[4]:

$$Q_A = \frac{1}{(ka)^3} + \frac{1}{ka} \tag{S-4}$$

Where $Q_A$ is the radiation quality factor, $k$ is the wave number and $a$ is the enclosing sphere centered on the antenna. With a dimension of 38×12×5.8 mm³, $ka$ can be calculated as[4]:

$$a = \sqrt{(38/2)^2 + (12/2)^2 + (5.8/2)^2} = 20.1 \text{ mm}$$

$$ka = \frac{f}{c} a = 2.5 \times 10^{-6}$$

The radiation quality factor of the ME antenna can be estimated as[4]:

$$Q_A = \frac{1}{(ka)^3} + \frac{1}{ka} = 6.46 \times 10^{16}$$

The quality factor of ME antenna can be calculated as[5]:

$$Q_{ME} = \frac{f_c}{BW_{3dB}} = 25.4$$

Based on Figure 2(b), the calculated $Q_{ME}$ is 25.4.

## 4. Impact of angle and position on the performance of the ME antenna

We further tested the impact of position and angle on the performance of the ME antenna. Figure S-4 shows experimental result of the normalized output voltage versus rotation angle around the center axis. Zhang *et al.* investigated the distribution of the magnetic field intensity of a planar rectangular coil. The research demonstrated that at the center of the coil, the magnetic field was perpendicular to the coil plane and the field intensity was independent of the rotation angle around the center z axis[6]. Therefore, the rotating of antenna will not affect the value of magnetic field strength, which is in accord with our experiment result. In this experiment, the output voltage only varies 2% at different angles, which is believed to due to the measurement error.

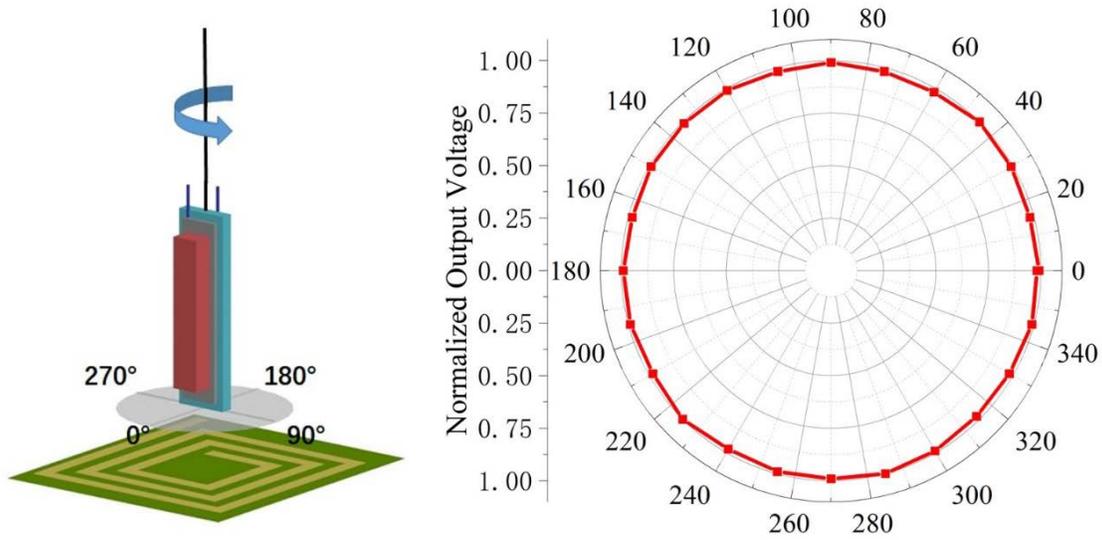

**Figure S-4** Experimental result of the normalized output voltage versus rotation angle around the center axis. The distance between transmitting antenna and receiving antenna is 5mm.

To further characterize the performance when the receiving antenna and transmitting spiral coil are in parallel, we change their relative location and record the output voltage, as illustrated in Figure S-5. The result reveals that when one end of the receiving antenna is at the center of the transmitting coil, the output voltage reaches a maximum. When the centers of the transmitting and receiving antennas are aligned at the same position (at 0 mm), the output voltage decreases significantly to only 4.5% of the maximum voltage. When receiving antenna is in parallel with the transmitting antenna, the traces on either side of the coil generate a magnetic field in parallel with the coil plane. The two magnetic field have the same intensity but are opposite direction. Therefore, when receiving antenna is across the two sides of coil, the strain in the Terfenol-D generated by the magnetic field by either side of the coil will have opposite direction, which reduces the output voltage. When the receiving antenna and the transmitting coil are aligned at the center, the strain generated by either side of coil has the same magnitude while opposite direction, which makes the output voltage to be almost zero. When the entire receiving antenna moves to one side of transmitting coil, the output voltage reaches a maximum since the strain generated on the receiving antenna is dominated by that side of coil. When the distance gradually increases to more than 17.5 mm, the distance between the receiving antenna and the spiral coil increases, which reduces the strain by the magnetic field generated by the spiral coil. As a result, the output voltage gradually decreases to zero. The maximum voltage occurs at $\pm 17.5$ mm.

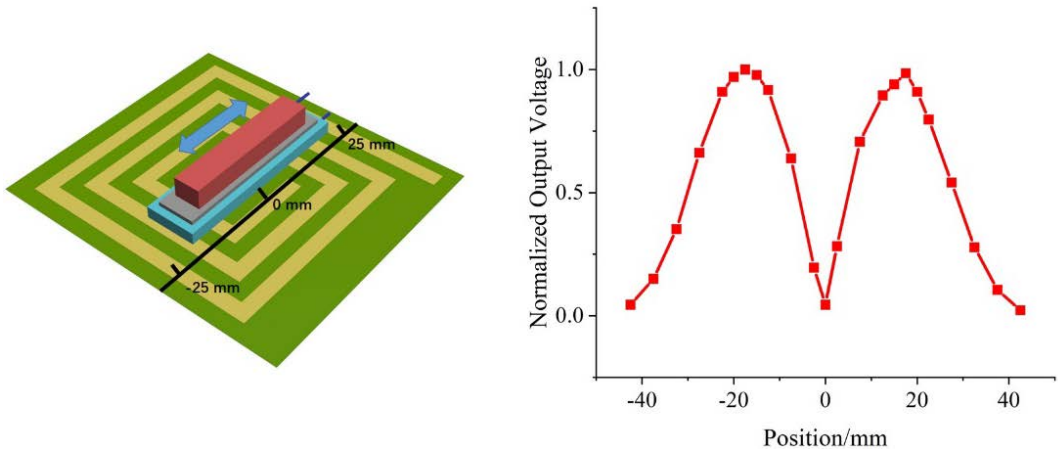

**Figure S-5** Experimental result of the receiving antenna translating horizontally in parallel with the transmitting coil. When receiving antenna is parallel to the transmitting coil, the maximum voltage occurs when one end of receiving antenna is at the middle of the transmitting coil.

**References of the supplementary materials**